\documentclass[aps,prd,preprint,superscriptaddress,tightenlines,nofootinbib,showpacs]{revtex4}

\usepackage{graphicx}
\usepackage{dcolumn}
\usepackage{bm}

\def\Y{\ifmmode \Upsilon \else%
$\Upsilon$ %
\fi}
\def\chib{\ifmmode \chi_b \else%
$\chi_b$ %
\fi}
\def\chibp{\ifmmode \chi_b' \else%
$\chi_b$ %
\fi}
\def\Q#1#2#3#4{\ifmmode
 \,#1\,{^{#2}#3}_{#4}
\else%
$#1\,{^{#2}#3}_{#4}$ %
\fi}
\def\eonem#1#2{\ifmmode
\left| <#1|r|#2> \right|
\else%
$\left| <#1|r|#2> \right|$
\fi}
\def\bb{b\bar b}

\def\ee{\ifmmode e^+e^- \else $e^+e^-$  \fi}
\def\mm{\ifmmode \mu^+\mu^- \else $\mu^+\mu^-$  \fi}
\def\LL{\ifmmode l^+l^- \else $l^+l^-$  \fi}

\def\etal{{\it et al.}}
\def\B{{\cal B}}
\def\chid{\chi^2_{1D}}

\begin{document}

\preprint{CLNS 04/1866}       
\preprint{CLEO 04-4}         

\title{First Observation of a $\Y(1D)$ State}


\author{G.~Bonvicini}
\author{D.~Cinabro}
\author{M.~Dubrovin}
\affiliation{Wayne State University, Detroit, Michigan 48202}
\author{A.~Bornheim}
\author{E.~Lipeles}
\author{S.~P.~Pappas}
\author{A.~Shapiro}
\author{A.~J.~Weinstein}
\affiliation{California Institute of Technology, Pasadena, California 91125}
\author{R.~A.~Briere}
\author{G.~P.~Chen}
\author{T.~Ferguson}
\author{G.~Tatishvili}
\author{H.~Vogel}
\author{M.~E.~Watkins}
\affiliation{Carnegie Mellon University, Pittsburgh, Pennsylvania 15213}
% ...
\collaboration{CLEO Collaboration} 
\noaffiliation


\date{\today}

\begin{abstract} 
We present the first evidence for the production of 
$\Y(1D)$ states in the four-photon cascade,
$\Y(3S)\to\gamma\chi_b(2P)$, 
$\chi_b(2P)\to\gamma\Y(1D)$, 
$\Y(1D)\to\gamma\chi_b(1P)$, 
$\chi_b(1P)\to\gamma\Y(1S)$,
followed by the $\Y(1S)$ annihilation
into $e^+e^-$ or $\mu^+\mu^-$.
The signal has a significance of  $10.2$ standard deviations.
The measured product branching ratio for these five decays,
$(2.5\pm0.5\pm0.5)\cdot 10^{-5}$, is consistent with the
theoretical estimates.
The data are dominated by the production of one 
$\Y(1D)$ state consistent with the $J=2$ assignment.
Its mass is determined to be 
$(10161.1\pm0.6\pm1.6)$ MeV, which is
consistent with the predictions from potential models and
lattice QCD calculations.

We also searched for $\Y(3S)\to\gamma\chib(2P)$, 
$\chib(2P)\to\gamma\Y(1D)$, followed by either 
$\Y(1D)\to\eta\Y(1S)$ or 
$\Y(1D)\to\pi^+\pi^-\Y(1S)$.
We find no evidence for such decays and
set upper limits on the product branching ratios.
\end{abstract}

\maketitle

Long-lived $b\bar b$ states are especially well suited for
testing lattice QCD  calculations \cite{LatticeQCD}
and effective theories
of strong interactions, 
such as potential models \cite{PotentialModels}
or NRQCD \cite{NRQCD}.
The narrow triplet-$S$ states, $\Y(1S)$, $\Y(2S)$ and
$\Y(3S)$, were discovered in 1977 in proton-nucleus
collisions at Fermilab \cite{UpsilonDiscovery}. 
Later, they were better resolved and studied at various
$e^+e^-$ storage rings.  
Six triplet-$P$ states, $\chi_b(2P_J)$ and $\chi_b(1P_J)$ with
$J=2,1,0$, were discovered in radiative
decays of the $\Y(3S)$ and $\Y(2S)$ 
in 1982 \cite{chib2pdiscovery} 
and  1983 \cite{chib1pdiscovery}, respectively.
There have been no observations of new narrow $\bb$ states since
then, despite the large number of such states predicted below the
open flavor threshold.

\begin{figure}
\includegraphics*[width=3.75in]{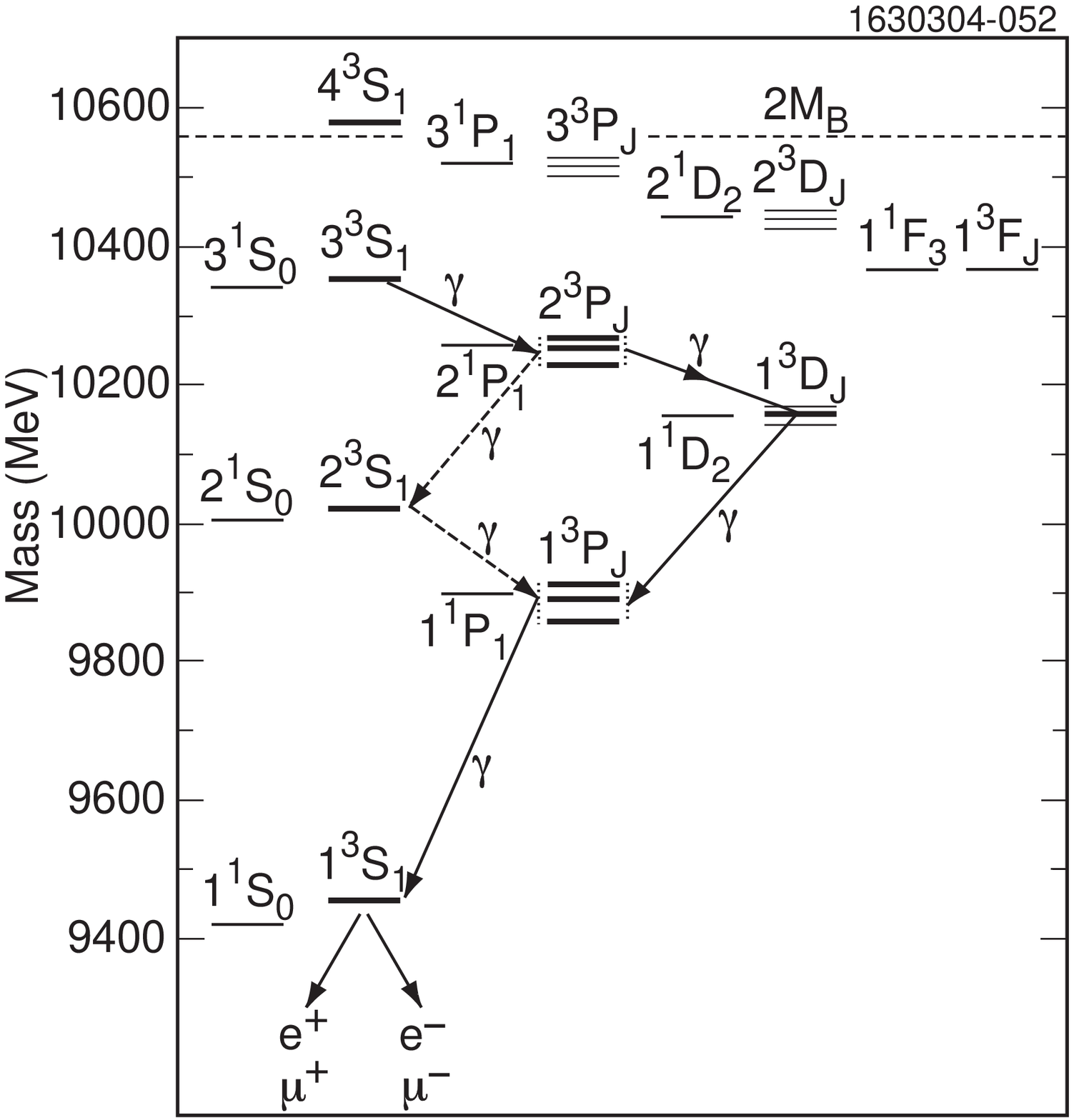}
\caption{The expected $\bb$ mass levels.
The four-photon transition sequence from the $\Y(3S)$ to the $\Y(1S)$
via the $\Y(1D)$ states is shown (solid lines).
An alternative route for the four-photon cascade via the $\Y(2S)$ state
is also displayed (dashed lines).}
\label{fig:level}
\end{figure}

In this paper, we present the first observation of the 
$\Y(1D)$ states.
They are produced in a two-photon
cascade starting from the $\Y(3S)$ resonance:
$\Y(3S)\to\gamma\chi_b(2P_J)$, 
$\chi_b(2P_J)\to\gamma\Y(1D)$. 
To suppress photon backgrounds from $\pi^0$s, which are copiously
produced in gluonic annihilation of the $b\bar b$ states,
we select events with two more subsequent photon transitions,
$\Y(1D)\to\gamma\chi_b(1P_J)$, 
$\chi_b(1P_J)\to\gamma\Y(1S)$, followed
by the $\Y(1S)$ annihilation into either 
$e^+e^-$ or $\mu^+\mu^-$
(see Fig.~\ref{fig:level}).
The product branching ratio for these five decays 
summing over $\Y(1D_{1,2,3})$ contributions was
predicted by Godfrey and Rosner \cite{GodfreyRosner}
to be $3.76\cdot 10^{-5}$.

The data set consists of 
$5.8\cdot 10^6$ $\Y(3S)$ decays observed with the CLEO III detector 
at the Cornell Electron Storage Ring (CESR). 
Charged particle tracking is done by a 47-layer
drift chamber and a four-layer silicon tracker which reside in a 1.5T solenoidal magnetic
field \cite{CLEOIIIDR}. 
Photons are detected using an electromagnetic calorimeter consisting of 
about 8000 CsI(Tl) crystals \cite{CLEOII}.
The particle-identification capabilities of
the CLEO III detector \cite{RICH}
are not used in the present analysis.

We select events with exactly four photons and two oppositely
charged leptons. The leptons must have momenta of at least 3.75 GeV.
We distinguish between electrons and muons by their 
energy deposition in the calorimeter. Electrons must have a high
ratio of energy observed in the calorimeter to the momentum measured
in the tracking system ($E/p>0.7$).
Muons are identified as minimum ionizing particles, and required to
leave $150-550$ MeV of energy in the calorimeter.
Stricter muon identification does not reduce background
      in the final sample, since all significant background
      sources contain muons.
Each photon must have at least 60 MeV of energy. We also ignore
all photons below 180 MeV in the calorimeter region closest to the
beam because of the spurious photons generated by 
beam-related backgrounds.
The total momentum of all photons and leptons in each event
must be balanced to within 300 MeV.
The invariant mass of the two leptons must be consistent 
with the $\Y(1S)$ mass within $\pm300$ MeV.

Much better identification of the $\Y(1S)$ resonance is
obtained by measuring  the mass of the system recoiling
against the four photons. 
The average resolution of the recoil mass is 17 MeV. 
The measured recoil mass is required to be within $-4$ and $+3$ standard
deviations from the $\Y(1S)$ mass.
The mass resolution of the produced $\Y(1D)$ state
depends on the measurement of the energies of the two lowest energy photons
in the event. 
Thus, we require that at least one of them is detected in the barrel 
part of the calorimeter, where the energy resolution is best.
The selected events are dominated at this point
by $\Y(3S)\to\pi^0\pi^0\Y(1S)$ transitions, which
have a branching ratio 
an order of magnitude higher than the expected signal
rate. In fact, the branching ratio measured for
a subsample of events in which two $\pi^0$ candidates
can be formed is consistent with the previous
measurements \cite{PDG}.
To suppress this background, we require the invariant mass
for any photon pair to be at least   
2 standard deviations away from the nominal $\pi^0$ mass.
The sum of such two $\pi^0$ mass deviations squared must be 
larger than 6 for any pairing scheme.

To look for $\Y(1D)$ events, we constrain events to be
consistent with a photon cascade from the $\Y(3S)$ to 
the $\Y(1S)$ via one of the $\chi_b(2P_J)$ and
one of the $\chi_b(1P_J)$ states.
Only $J=1$ or $2$ are used since the $J=0$ states
have small decay fractions for electromagnetic transitions.
For each $J_{2P}$, $J_{1P}$ combination we calculate a chi-squared:
$$
\chi^2_{1D,\,J_{2P},\, J_{1P}}(M_{\Y(1D)})=\sum_{j=1}^4 
\left( \frac{E_{\gamma\,j}-E_{\gamma\,j}^{expected}(M_{\Y(1D)},\,J_{2P},\, J_{1P})}{
             \sigma_{E_{\gamma\,j}} } \right)^2,
$$
where $E_{\gamma\,j}$ are the measured photon energies;
$E_{\gamma\,j}^{expected}$ are the expected photon energies calculated from
the known masses of the $b\bar b$ states and the measured photon directions
in each event. The masses of the $\Y(1D)$ states are 
not known. Therefore, we minimize the above chi-squared with respect
to $M_{\Y(1D)}$ which is allowed to vary for each event.
The above formalism requires that we know how to order the four
photons in the cascade. While the highest energy photon must be due to the
fourth transition, and the second highest energy photon must be
due to the third transition, there is sometimes an ambiguity
in the assignment of the two lower energy photons from the
first two transitions, since the range of photon energies
in the $\Y(3S)\to\gamma\chi_b(2P_J)$ decay overlaps the similar
energy range in the $\chi_b(2P_J)\to\gamma\Y(1D)$
transition. We choose the combination that minimizes 
the above chi-squared.
There are four possible combinations of  $J_{2P}$, $J_{1P}$ values. 
We try all of them and choose the one that produces
the smallest chi-squared, 
$\chi^2_{1D}=min\,\chi^2_{1D,\,J_{2P},\, J_{1P}}$.

In addition to the four-photon cascade via the $\Y(1D)$ states, our
data contain events with the four-photon cascade via the $\Y(2S)$ state:
$\Y(3S)\to\gamma\chi_b(2P_J)$, 
$\chi_b(2P_J)\to\gamma\Y(2S)$, 
$\Y(2S)\to\gamma\chi_b(1P_J)$, 
$\chi_b(1P_J)\to\gamma\Y(1S)$,
$\Y(1S)\to l^+l^-$
(see Fig.~\ref{fig:level}).
The product branching ratio for this entire decay sequence
(including $\Y(1S)\to l^+l^-$)
is predicted by Godfrey and Rosner \cite{GodfreyRosner}
to be $3.84\cdot10^{-5}$, thus
comparable to the predicted $\Y(1D)$ production rate.
In these events, the second highest energy photon
is due to the second photon transition
(see Fig.~\ref{fig:level}).
Unfortunately, these events can sometimes be confused
with the $\Y(1D)$ events due to our limited 
experimental energy resolution. The second and
third photon transitions in the $\Y(2S)$ cascade sequence
can be mistaken for the third and second transitions
in the $\Y(1D)$ cascade sequence, respectively.
Therefore, it is important to suppress the $\Y(2S)$ cascades.
We achieve this by finding the $J_{2P}$, $J_{1P}$ ($=0,1$ or $2$) 
combination that minimizes the associated chi-squared 
for the $\Y(2S)$ hypothesis,
$\chi^2_{2S}=min\,\chi^2_{2S,\,J_{2P},\, J_{1P}}$,
where $\chi^2_{2S}$ is exactly analogous to $\chi^2_{1D}$ with
the $M_{\Y(1D)}$ replaced with $M_{\Y(2S)}$. 
We then require $\chi^2_{2S}>12$.
Notice that the masses of all intermediate states are known for
the $\Y(2S)$ cascade, thus this variable is more
constraining than $\chi^2_{1D}$. 

To further suppress the $\Y(2S)$ cascade  events, we construct a
quasi-chi-squared variable, $\chi^{2+}_{2S}$, that sums in quadrature
only positive deviations of the measured photon energies
from their expected values. This variable is less sensitive than $\chi^2_{2S}$
to fluctuations in the longitudinal and transverse energy leakage 
in photon showers that sometimes produce large negative energy deviations
and correspondingly a large $\chi^2_{2S}$ value.
With the additional criteria $\chi^{2+}_{2S}>3$
and $\chi^2_{1D}<10$, the cross-feed efficiency
for $\Y(2S)$ events is reduced to 0.3\%, while
the signal efficiency is 12\%.
The $\pi^0\pi^0$ background cross-feed efficiency is 
0.02\%. 
Monte Carlo simulation of the signal events is based
on the photon transition rate predicted for the $J=2$
$\Y(1D)$ state by Godfrey and Rosner \cite{GodfreyRosner}.
We use the $J=1$ assumption to estimate the model dependence of
the signal efficiency. 
The proper angular distribution of the first photon in the cascade,
$\Y(3S)\to\gamma\chi_b(2P)$, is taken into account,
resulting in a 4\%\ relative change of the efficiency 
compared to the uniform distribution.
Angular correlations in the subsequent photon transitions
are neglected.

\begin{figure}
\includegraphics*[width=3.75in]{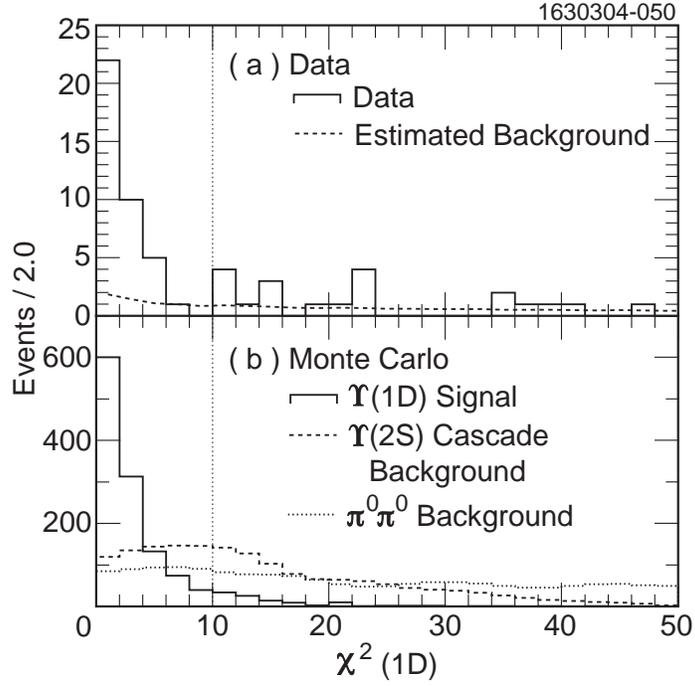}
\caption{Distributions of $\chid$ for (a) data and (b) Monte Carlo simulations of
the signal and backgrounds.
The solid histogram in (a) represents the data, while the dashed line
represents the background fit described in the text.
The solid histogram in (b) represents the $\Y(1D_2)$ signal Monte Carlo.
The dashed histogram shows the simulated background 
from the $\Y(2S)$ cascades. 
This distribution is scaled up by a factor of 10 
in efficiency normalization
to make it visible
when superimposed on that of the signal Monte Carlo. 
The dotted histogram shows the Monte Carlo 
distribution for $\pi^0\pi^0$ transitions with the $\pi^0$ cuts removed,
normalized to the number of entries in the $\Y(2S)$
cascade background histogram.
The vertical line indicates the cut value used for the $\Y(1D)$ mass 
analysis.
}
\label{fig:chid}
\end{figure}

The data $\chi^2_{1D}$ distribution 
after all these cuts is shown by the solid histogram in 
Fig.~\ref{fig:chid}a. 
A narrow peak near zero is observed,
just as expected for $\Y(1D)$ events.
The signal Monte Carlo distribution for $\Y(1D)$ events is shown 
by the solid histogram
in Fig.~\ref{fig:chid}b. 
The background Monte Carlo distribution for the $\Y(2S)$ cascades, 
after a factor of 10 enhancement relative to the 
$\Y(1D)$ normalization, is also shown for comparison.
The $\Y(3S)\to\pi^0\pi^0\Y(1S)$ Monte Carlo distribution is shown without
the $\pi^0$ veto cuts to increase the statistics.
We conclude that the backgrounds cannot produce
as narrow a peak as observed in the data. 

After all the selection cuts,
we observe 38 events in the data with $\chid<10$. 
The background estimates are
$1.5\pm1.4$ and $1.3\pm0.9$ $\Y(2S)$ and $\pi^0\pi^0$ 
events, respectively. 
The errors on the background estimates include
systematic effects.
Feed-across from the other photon and hadronic 
transitions is found to be negligible.
Continuum backgrounds, for example
due to radiative Bhabha scattering events, were
estimated to contribute $0.7\pm0.7$ events,
using data taken at the $\Y(1S)$ 
resonance.
After the background subtractions, the estimated
signal yield is $34.5\pm6.4$ events.
 
An alternative background subtraction method is
obtained by fitting the $\chi^2_{1D}$ distribution
in the range between 0 and 100
to the Monte Carlo predicted
signal and background
contributions. In this method the background normalization
is effectively determined by the event yield observed
in the tail of the $\chi^2_{1D}$ distribution.
The background shape is 
assumed to follow the $\pi^0\pi^0$ Monte Carlo distribution
with the $\pi^0$ veto cuts removed to increase the
Monte Carlo statistics (see Fig.~\ref{fig:chid}b).  
A linear background fit was also
tried and yielded similar results.
The $\Y(2S)$ background is fixed
in this fit to the Monte Carlo simulation,
 normalized to the rate predicted 
by Godfrey and Rosner, 
since unlike all other backgrounds it tends to
peak near the signal region.
This method yields $38.5\pm6.8$ signal events with a signal
efficiency of $13\%$ in the extended $\chi^2_{1D}$ range.

The significance of the signal is evaluated from the change of 
likelihood between the nominal fit and when fitting the data 
with the background shapes alone 
and corresponds to
10.2 standard deviations ($8.9\sigma$ for $\gamma\gamma\gamma\gamma\mm$
and $5.1\sigma$ for $\gamma\gamma\gamma\gamma\ee$).
The signal product branching ratio
obtained with both methods of background subtraction 
is the same, 
$\B(\gamma\gamma\gamma\gamma\LL)_{\Y(1D)}
=(2.5\pm0.5\pm0.5)\cdot10^{-5}$.
Throughout this paper we quote branching ratios 
averaged over the $\mm$ and
$\ee$ channels.
The first error is statistical, while the second error is
systematic. The systematic error includes uncertainty in
the background subtraction (8\%), model dependence of
the efficiency (8\%), uncertainty in the detector simulation (8\%)
and the number of $\Y(3S)$ decays (2\%).
This branching ratio is consistent with the theoretically 
estimated rate \cite{GodfreyRosner}.

\begin{figure}
\includegraphics*[width=3.75in]{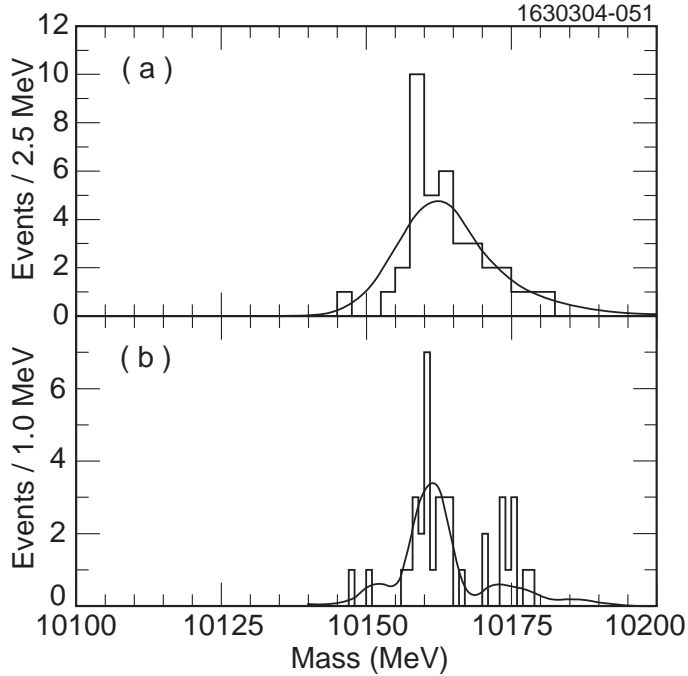}
\caption{Distributions of the measured $\Y(1D)$ mass in the data using 
(a) the recoil mass method, and 
(b) the $\chid$ fit method.
The results of fits for a single $\Y(1D)$ state are superimposed.
The $\chid$ fit method produces satellite peaks as
explained in the text.
}
\label{fig:mass}
\end{figure}

A straightforward way to measure the mass of the produced $\Y(1D)$ state 
is to calculate the mass of the system recoiling against the two lower
energy photons in the event.
This distribution is shown in Fig.~\ref{fig:mass}a.
The width of the observed peak is consistent with the detector
resolution, implying the data are dominated by production of just
one $\Y(1D)$ state. 
We use the signal line shape obtained from
the Monte Carlo simulations to fit the data 
and determine the mass of this state.

Another estimate of the true $\Y(1D)$ mass is given by
the mass value that minimizes $\chi^2_{1D}$.
 This distribution is shown in
Fig.~\ref{fig:mass}b. 
The data are again consistent with
the single-peak hypothesis. 
The fit to the
expected signal shape from Monte Carlo simulations is superimposed in
the figure.
While this method has a
mass resolution of about 3 MeV, compared to a value of about 7 MeV
for the missing-mass technique, 
the signal shape here has a complicated
tail structure originating from photon 
energy fluctuations which
can make a wrong $J_{2P}$, $J_{1P}$ combination produce the
smallest chi-squared value. This produces
small satellite peaks on both sides of the main peak.
Both methods of mass determination give consistent results.
The mass of the observed state is determined to be
$(10161.1\pm0.6\pm1.6)$ MeV, where the first error is
statistical and the second systematic. The systematic
error includes the measurement method dependence ($\pm1.2$ MeV)
and the mass calibration error ($\pm1.1$ MeV).
The significance of a possible second peak 
around $10175$ MeV is only 1.9 standard
deviations. The recoil mass distribution
discussed in the previous paragraph 
and shown in Fig.~\ref{fig:mass}a 
favors no events in any second peak.
Thus, we cannot claim observation of more than
one $\Y(1D)$ state based on our data.

The $\chid$ minimization favors 
the $J_{2P}=1$, $J_{1P}=1$ cascade path for
most of the observed events, 
indicating that the observed state is
either $J_{1D}=1$ or $2$. Theoretically, the production
rate of the $J_{1D}=2$ state is expected to be 6 times
larger than for the $J_{1D}=1$ state \cite{GodfreyRosner}.
Therefore, we
interpret our signal as coming predominantly
from the production of the $\Upsilon(1^3D_2)$.
Small contributions of the $\Upsilon(1^3D_1)$ and 
$\Upsilon(1^3D_3)$ with masses close to the
observed $\Upsilon(1^3D_2)$ mass cannot be
ruled out. However, they are impossible to
quantify from our data alone without
prior knowledge of the fine-structure mass splitting.

The measured mass is in good agreement with the mass
of the $\Y(1^3D_2)$ state predicted by 
lattice QCD calculations \cite{LatticeQCD}
and those potential
models which also give a good fit to the other known
$\bb$ states \cite{lp03-skwa}.
All potential model calculations predict
the $\Upsilon(1^3D_2)$ mass to be between
$0.5$ and $1.0$ MeV lower than the 
center-of-gravity (c.o.g.) mass for this triplet.
Adding this theoretical input to our results, we
obtain $(10162\pm2)$ MeV for the c.o.g.\
mass, where we assigned an additional uncertainty of
1 MeV to the correction for the $1^3D_2-$c.o.g.\ mass difference.

\begin{figure}
\includegraphics*[width=3.75in]{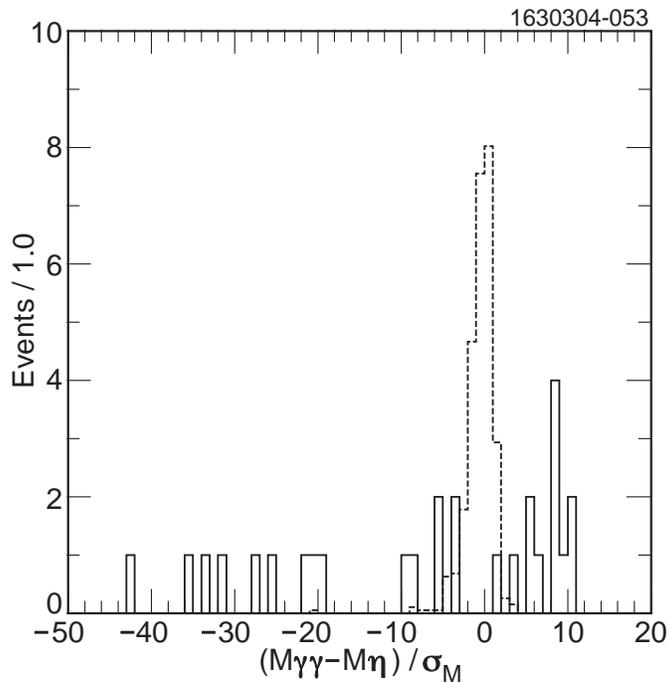}  
\caption{Distribution of the deviation of the two-photon mass from
the $\eta$ mass divided by the estimated mass resolution for
$\Y(1D)\to\eta\Y(1S)$ candidates from the data (solid histogram)
and from the signal Monte Carlo simulation (dashed histogram).}  
\label{fig:ggetadata}
\end{figure}

Voloshin recently suggested  that the 
$\eta$ transition could be enhanced in
$\Y(1D)\to\Y(1S)$ decays \cite{VoloshinEta}.
Since the $\eta$ often decays to two photons,
we can look for it in the same sample
preselected for the four-photon cascade
analysis. 
We reverse the $\chid$ cut
($\chid>10$) to suppress the 
four-photon cascades via the $\Y(1D)$
states. Otherwise they would contribute a smooth 
background to our $\eta$ search variable
(defined below). Since we still want  
the two-photon cascade to produce a $D-$state
via $\Y(3S)\to\gamma\chi_b(2P_{2,1})$,
$\chi_b(2P_{2,1})\to\gamma\Y(1D)$,
we require that one of the two lowest energy 
photons fits the $\Y(3S)\to\gamma\chi_b(2P_{2,1})$
transition ($70.0<E_\gamma<110$ MeV).
Because the backgrounds are small, we
did not constrain the second photon
energy and therefore we did not restrict 
the sample to
any particular value of $\Y(1D)$ mass.
The signal efficiency is 13\%
(not including $B(\eta\to\gamma\gamma)$).
To search for the eta we analyze the invariant 
mass distribution for the 
two most energetic photons.
The distribution of 
$(M_{\gamma\gamma}-M_\eta)/\sigma_M$
for the data is shown in
Fig.~\ref{fig:ggetadata},
where $\sigma_M$ is the expected
$\eta$ mass resolution. 
No signal is observed. 
  To estimate the upper limit we fit this distribution with the 
  eta line shape and a smooth approximation for the background 
  obtained from the Monte Carlo simulations.
The corresponding 90\% C.L.
upper limit on the
product branching ratio is:
$\B(\Y(3S)\to\gamma\gamma\Y(1D))
 \B(\Y(1D)\to\eta\Y(1S))
 \B(\Y(1S)\to\LL)<0.6\cdot10^{-5}$
or
$\B(\Y(3S)\to\gamma\gamma\Y(1D))
 \B(\Y(1D)\to\eta\Y(1S))<2.3\cdot10^{-4}$
if we use the world average value for
$\B(\Y(1S)\to\LL)$ \cite{PDG}.
A systematic error of 8.3\% is included
by scaling up the upper limit by one unit 
of the systematic error.
Dividing the estimated
upper limit by the measured 
product branching ratio 
for the four-photon cascade,
we obtain:
$\B(\Y(1D)\to\eta\Y(1S))/
 \B(\Y(1D)\to\gamma\gamma\Y(1S))
 < 0.25 $ (at 90\%\ C.L.).
Common systematic errors
were taken out in this calculation.

\begin{figure}
\includegraphics*[width=3.75in]{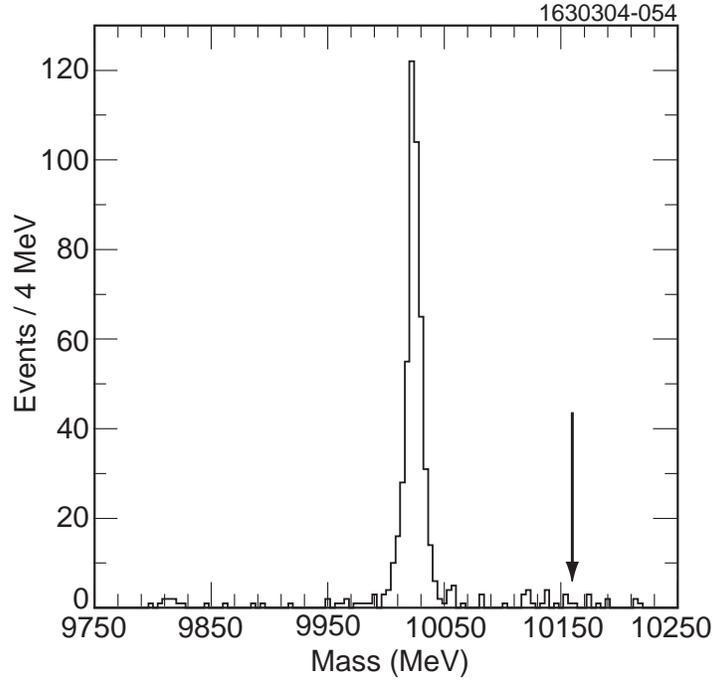} 
\caption{
The invariant mass distribution for the system 
recoiling against the two photons in
$\Y(3S)\to\gamma\gamma\pi^+\pi^-\Y(1S)$ events.
The observed peak is due to transitions via the $\Y(2S)$ state,
followed by $\Y(2S)\to\pi^+\pi^-\Y(1S)$.
The arrow indicates where the signal due to transitions via
the $\Y(1^3D_2)$ state is expected.}
\label{fig:ggppmdata}
\end{figure}

Predictions for the branching ratio of $\Y(1D)\to\pi^+\pi^-\Y(1S)$
vary by orders of magnitude among various theoretical
predictions (from $0.2\%$ to $49\%$) \cite{1dpipi}.
To look for these transitions, we 
selected $\gamma\gamma\pi^+\pi^-\LL$ events
using similar selection cuts to our
$\gamma\gamma\gamma\gamma\LL$ analysis.
After requiring the di-lepton mass and 
the recoil mass against the $\gamma\gamma\pi^+\pi^-$
 to be consistent with
the $\Y(1S)$ mass, and checking that the total momentum of
the event is consistent with zero, 
we require at least one photon to have an energy
in the $70-110$ MeV range, corresponding to 
the $\Y(3S)\to\gamma \chi_b(2P_{2,1})$
transition.
We then 
measure the mass of the intermediate $b\bar b$ state,
assuming that it is produced by the two-photon
cascade. 
This mass can be estimated by using either
the photons or the pions. 
To get the best estimate, we
average the two mass estimates
by giving them weights inversely proportional 
to the mass resolution squared, as determined
by Monte Carlo simulations. 
The weights are 40\%\ for the
$\gamma\gamma$ recoil mass, and 60\%\
for the mass obtained using $\pi^+\pi^-$.
The signal efficiency is 19\%.
The resulting mass distribution is
shown in 
Fig.~\ref{fig:ggppmdata}.
The prominent peak observed in the data
is due to 
$\Y(3S)\to\gamma\chi_b(2P)$,
$\chi_b(2P)\to\gamma\Y(2S)$,
$\Y(2S)\to\pi^+\pi^-\Y(1S)$.
From a fit to this peak, we 
determine the product branching
ratio for this $\Y(2S)$ decay signal to be
$1.13\pm0.16$ times the value
derived from the individually
measured transition rates \cite{PDG}.
This provides a good check for our
detection efficiency.

There is no indication of any
excess of events 
at the $\Y(1D)$ mass value observed
in our four-photon cascade analysis.
To estimate an upper limit on the
signal rate, we fit the data
with a signal fixed at our observed $\Y(1^3D_2)$
mass and a smooth background parameterized by 
a cubic polynomial.
The following limits (90\%\ C.L.) are
obtained:
$\B(\Y(3S)\to\gamma\gamma\Y(1D_2))
 \B(\Y(1D_2)\to\pi^+\pi^-\Y(1S))
 \B(\Y(1S)\to\LL)<2.7\cdot10^{-6}$
or
$\B(\Y(3S)\to\gamma\gamma\Y(1D_2))
 \B(\Y(1D_2)\to\pi^+\pi^-\Y(1S))
 <1.1\cdot10^{-4}$.
Dividing our upper limit by
the measured rate for the four-photon
cascade we obtain:
$\B(\Y(1D_2)\to\pi^+\pi^-\Y(1S))/
 \B(\Y(1D_2)\to\gamma\gamma\Y(1S))
 < 1.2$ (at 90\%\ C.L.).
We also set an upper limit for the
production of any $\Y(1D)$ state 
(followed by $\pi^+\pi^-\Y(1S)$ decay)
with
a mass in the $10140-10180$ MeV range, which
comfortably covers the predicted size of fine-structure 
splitting for the $\Y(1D)$ triplet \cite{GodfreyRosner}.
Here, we do not try to subtract backgrounds
and accept all 9 events observed
in this mass range as signal candidates.
This results in the following upper limits:
$\B(\Y(3S)\to\gamma\gamma\Y(1D))
 \B(\Y(1D_J)\to\pi^+\pi^-\Y(1S))
 \B(\Y(1S)\to\LL)<6.6\cdot10^{-6}$
or 
$\B(\Y(3S)\to\gamma\gamma\Y(1D))
 \B(\Y(1D_J)\to\pi^+\pi^-\Y(1S))
 <2.7\cdot10^{-4}$
for a sum over all different $J_{1D}$ values.

These upper limits are inconsistent
(lower by a factor of about 7)
with the rate estimated by 
Rosner \cite{1dpipi} using
the Kuang-Yan model for 
$\Gamma(\Y(1D)\to\pi^+\pi^-\Y(1S))$ \cite{1dpipi-KY}
and a factor of about 3 higher than the predicted rate
based on the model by 
Ko \cite{1dpipi-K}. 
Our upper limits are about 30 times 
higher than those predicted by Moxhay's model \cite{1dpipi-M}.

In summary, we present
the first significant evidence for the production of the 
$\Y(1D)$ states in the four-photon cascade
$\Y(3S)\to\chi_b(2P)\to\Y(1D)\to\chi_b(1P)\to\Y(1S)$.
The data are dominated by the production of one 
$\Y(1D)$ state, consistent with the $J=2$ assignment.
Its mass is determined to be 
$(10161.1\pm0.6\pm1.6)$ MeV,
in agreement with the potential models and lattice
QCD calculations.
The measured product branching ratio, 
$(2.5\pm0.5\pm0.5)\cdot 10^{-5}$,
is consistent with the
theoretical estimate, especially when
comparing with the predicted rate for
the $\Y(1D_2)$ state alone, $2.6\cdot 10^{-5}$ 
\cite{GodfreyRosner}.

We have also searched for $\Y(3S)\to\gamma\chib(2P)$, 
$\chib(2P)\to\gamma\Y(1D)$ followed by either 
$\Y(1D)\to\eta\Y(1S)$ or 
$\Y(1D)\to\pi^+\pi^-\Y(1S)$.
We find no evidence for such decays and
set upper limits on the product branching ratios.
The latter are inconsistent with the Kuang-Yan model which
predicts a large $\Y(1D)\to\pi^+\pi^-\Y(1S)$ width. 

We gratefully acknowledge the effort of the CESR staff 
in providing us with
excellent luminosity and running conditions.
This work was supported by 
the National Science Foundation,
the U.S. Department of Energy,
the Research Corporation,
and the 
Texas Advanced Research Program.

\newpage

\end{document}